\documentclass[times]{TRR}

\usepackage{moreverb,url}
\usepackage{lscape}
\usepackage{adjustbox}
\usepackage{tablefootnote}
\usepackage{lipsum}
\usepackage{tabularx}
\usepackage{threeparttable}
\usepackage{natbib}
\usepackage{subfigure}
\usepackage{multirow}
\usepackage{lineno}

\usepackage[bookmarksopen,bookmarksnumbered,citecolor=red,urlcolor=red]{hyperref}

\newcommand\BibTeX{{\rmfamily B\kern-.05em \textsc{i\kern-.025em b}\kern-.08em
T\kern-.1667em\lower.7ex\hbox{E}\kern-.125emX}}

\def\volumeyear{2020}

\begin{document}

\runninghead{Zhou et.al.}

\title{Review of Learning-based Longitudinal Motion Planning for Autonomous Vehicles: Research Gaps between Self-driving and Traffic Congestion
}

\author{Hao Zhou\affilnum{1}, Jorge Laval\affilnum{1*}, Anye Zhou\affilnum{1}, Yu Wang\affilnum{1}, Wenchao Wu\affilnum{1}, Zhu Qing\affilnum{1} and Srinivas Peeta\affilnum{1}}

\affiliation{\affilnum{1} Civil and Environmental Engineering, Georgia Tech, US}

\corrauth{Jorge Laval, jorge.laval@ce.gatech.edu}

\begin{abstract}
Self-driving technology companies and the research community are accelerating their pace to use machine learning longitudinal motion planning (mMP) for autonomous vehicles (AVs). This paper reviews the current state of the art in mMP, with an exclusive focus on its impact on traffic congestion. We identify the availability of congestion scenarios in current datasets, and summarize the required features for training mMP. For learning methods, we survey the major methods in both imitation learning and non-imitation learning. We also highlight the emerging technologies adopted by some leading AV companies, e.g. Tesla, Waymo, and Comma.ai. We find that: i) the AV industry has been mostly focusing on the long tail problem related to safety and overlooked the impact on traffic congestion, ii) the current public self-driving datasets have not included enough congestion scenarios, and mostly lack the necessary input features/output labels to train mMP, and iii) albeit reinforcement learning (RL) approach can integrate congestion mitigation into the learning goal, the major mMP method adopted by industry is still behavior cloning (BC), whose capability to learn a congestion-mitigating mMP remains to be seen.
Based on the review, the study identifies the research gaps in current mMP development. Some suggestions towards congestion mitigation for future mMP studies are proposed: i) enrich data collection to facilitate the congestion learning, ii) incorporate non-imitation learning methods to combine traffic efficiency into a safety-oriented technical route, and iii) integrate domain knowledge from the traditional car following (CF) theory to improve the string stability of mMP.  
\end{abstract}

\maketitle

\section{Introduction}

Self-driving cars are around the corner, quite literally. And yet, despite numerous studies \citep{van2006impact,shladover2012impacts,talebpour2016influence,Mahmassani2016,Talebpour2017,kesting2008adaptive,delis2015macroscopic} on the potential impacts of AVs and \textcolor{black}{connected and autonomous vehicles (CAVs)} on  traffic flow,  a reliable car-following (CF) model describing the longitudinal dynamics of AVs is still lacking. \textcolor{black}{This makes evaluating their impact on traffic flow challenging. Recent empirical experiments reveal that the existing longitudinal control systems on level-2 AVs are string unstable \textcolor{black}{ \citep{gunter2019model,gunter2020commercially,li2021car}}, which indicates small perturbations (e.g., speed fluctuations) tend to grow upstream of a platoon, and eventually lead to full stop-and-go motions. Those empirical findings are surprising, which indicates the AVs might cause more traffic congestion even than human drivers. The results also distinguish from the successful design or implementation of the string stable longitudinal controller in the literature, including both of the adaptive cruise control (ACC) and cooperative adaptive cruise control (CACC) algorithms \citep{naus2010string,Naus2010,Ploeg2011,Bu2010,Zhou2020SmoothSwitchingCC}. We conjecture that the gap between the practice and the theory may result from: i) the longitudinal control of the level-2 AVs, aka ACC, does not factor in string stability in its design}; ii) in the real-world scenarios, some other issues (e.g., safety, efficiency, comfort or user acceptability) are weighted more than string stability performance, thus, the controller will suppress the string stability properties to satisfy other performance metrics \cite{zhou2021impact}; and iii) the hardware equipment (sensing devices and actuators) is not capable of realizing the string-stable control command. The rough and choppy measurements, and the slow-response actuator equipped on economy daily cars enforce the control command to be heavily filtered before exerted on the vehicle (otherwise the vehicle will behave in a undesired jerky manner), which makes string stability not achievable.
Given the undesired string unstable ACC, it is possible that current AV systems might induce more instability than human drivers, which could induce more traffic congestion and emissions. \textcolor{black}{From a traffic perspective, there is a critical need for a deeper understanding of AVs' longitudinal behaviors to predict their impact on traffic congestion}.



\begin{table*}[ht]
\caption{Latest ADAS technologies from major automakers in 2020}
\label{new_ACC}
\begin{tabularx}{\linewidth}{lXXX}
\hline
     \textbf{Automaker} & \textbf{Technology}  & \textbf{Sensors} & \textbf{Description}
     \\
     \hline
    Tesla & Autopilot and FSD & Radar + 8 cameras & Traffic-aware cruise control \citep{FSD2020}  \\ \hline
    Nissan & ProPILOT 2.0 & 7 cameras, 5 radars and 12 sonars & Incorporates 3D high definition map \citep{Nissan}\\ \hline
     Toyota & Full-speed range DRCC & Radar + camera & Work in full-speed range \citep{Toyota}\\ \hline
     Honda & Honda sensing &  Radar + camera &
     Collision mitigation braking, speed signs \citep{Honda}\\ \hline
     GM& ACC-camera & Camera+radar & ACC is based on camera\\
     \hline
    Ford & Intelligent ACC & Radar + camera & Automatically adapt to speed limit signs \citep{Ford}\\ \hline
    Audi & ACC with stop \& go & Two radars + camera & Stop-and-go \citep{audi}  \\ \hline
    BMW & ACC with stop \& go & Radar + camera & Stop \& go and speed limit compliance \citep{BMW} \\ 
     \hline
\end{tabularx}
\vspace{-6mm}
\end{table*}

\textcolor{black}{Meanwhile, the current AV technology is fast evolving thanks to the recent advancement in computer vision and machine learning.}
Notably, we are witnessing a fundamental shift from the traditional radar-based ACC, which solely relies on radar \citep{ajanovic2018search}, to the camera-included Advanced driver-assistance systems (ADAS). The transition is reasonable and as expected, because the traditional radar-based ACC has a limited functionality due to its pure reliance on the radar sensor and the hard-coded human-crafted rules. Additionally, the inherent structure of radar-based ACC may lead to issues such as: i) radar-based ACC cannot adapt to variable speed limits, respond to the ambient traffic proactively or predict the upcoming incidents, ii) traditional radar-based ACC cannot navigate in stop-and-go traffic due to the limitations in detecting slow-moving or still objects, and iii) in order to alleviate the traffic oscillations, the hard-coded CF rules also requires more human efforts in examining and tuning the controller. 

\textcolor{black}{This shift from radar to cameras can be game-changing because vision open the gate for incorporating more machine learning methods such as mMP for planning}. The leading company Tesla is famous for their camera-based autonomy solution and their latest full self-driving (FSD) function is featured by the 'traffic-aware cruise control' \citep{FSD2020}. \textcolor{black}{Starting from May 2021, Tesla has fully ditched radar on their new releases of the FSD softwares \citep{tesla-ditch}.}
Although FSD's cruise control demonstrates multiple intelligent features, there is no reliable evidence to show whether its longitudinal motion planning is powered by neural networks, or the traditional rule-based ACC with extra augmentations. Recently, many other automakers are catching up and also start to integrate cameras into the longitudinal control module. A brief summary could be seen in Table.\ref{new_ACC}. In general those automakers adopt a similar ADAS, which adds camera for lane-keeping, collision avoidance, and enables the low-speed cruise control in stop-and-go traffic where a single radar often fails. GM and Nissan seem to be slightly different. Instead of using radar, GM's current ACC function is reportedly only using camera \citep{GM}, and its upcoming Super Cruise \citep{Super-cruise} would be a hands-free function thanks to Lidar maps of highways. Nissan \citep{ProPilot} has delivered a level-3 autonomous driving using a complex suite of sensors similar to Tesla. Nissan also claims to be the first automaker that incorporates the 3D high-definition map.



On the other hand, although there exist hundreds of AV automakers, their AV service providers are far less. Here we \textcolor{black}{summarize the major ADAS} service providers with their major customers and collaborators in Fig.\ref{ADAS_major}. More detailed service providers of ADAS and other AV technologies are attached in the Appendix. \textcolor{black}{It indicates that despite the AVs have different brands, their impact on traffic flow can be similar to each other.}

\begin{figure}[!htbp]
    \includegraphics[width=0.5 \textwidth]{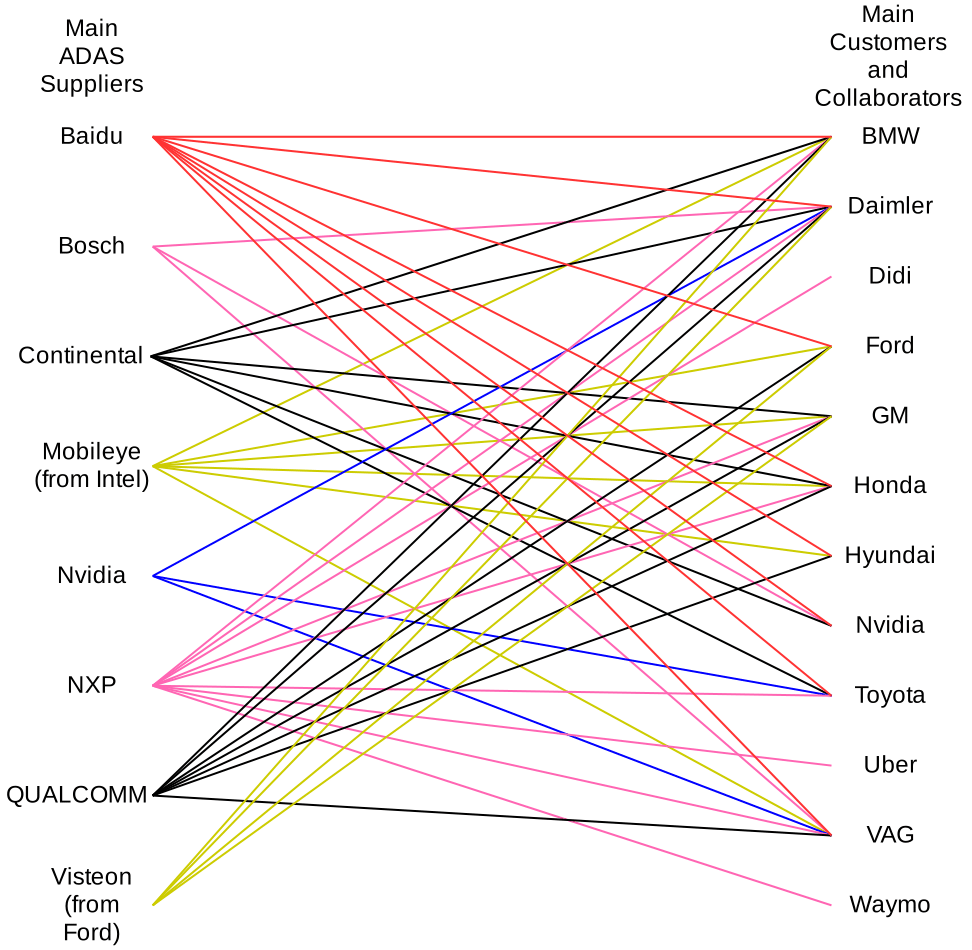}
    \centering
    \caption{Main suppliers and customers of ADAS. Source: \citep{Wenchao2020}}
    \label{ADAS_major}
\end{figure}

\textcolor{black}{While the level-2 market AVs are proprietary and we have no explicit knowledge about their longitudinal control methods,} the self-driving technology companies/institutions have been more transparent and exhibited a clear goal to achieve and adopt the mMP. Waymo published their feature-engineering mMP approach in \cite{bansal2018chauffeurnet}. Remarkably, an end-to-end mMP model was recently  open-sourced by Comma.ai, an aftermarket self-driving company which retrofits regular cars with a mono-camera phone. Similar self-driving service is also seen at Mobieye \citep{Mobileye} of Intel, which helps make regular car become AVs with only a single camera device.
Similarly, many other self-driving technology companies have published their datasets which indicate mMP methods towards the longitudinal autonomy. Readers are referred to \cite{scale} for a full list of those public self-driving datasets, which are filtered by data type, traffic scenario diversity, and annotation. On the other hand, a plethora of research papers \citep{,yu2017baidu,george2018imitation,sharma2019lateral,kuutti2019end,pathak2019adaptive} have been proposed to accomplish mMP using different learning approaches. 

\textcolor{black}{With all that being said, it is highly possible that mMP will be the future of AVs, for both of the level-2 commercial vehicles or the higher-level full self-driving cars according to the definition of SAE (Society of Automotive Engineers). 
As its impacts on traffic congestion are essential and have not received enough attention yet, it is necessary to provide an in-depth review to understand the state-of-the-art mMP methods, with the purpose of promoting more traffic-friendly AVs in the long run.}

\textcolor{black}{There already exists some review works on AV planning algorithms in literature, but for whom the focus is not related to traffic congestion or mMP methods}. For example, \cite{babak2017control} is limited to the engineering perspective only, focusing on sensors and embedded systems for AVs. \citep{katrakazas2015real,paden2016survey, ajanovic2018search} in the robotics literature discussed the traditional motion planning approaches like graph search, trajectory optimization \textcolor{black}{ and optimal control methods, which are out of scope for this study.} Quite a few reviews focused on the rule-based AV control, especially for CACC \cite{dey2015review}. 
Attempts to consolidate more relevant studies on mMP of AVs are available. \cite{ni2020survey} \textcolor{black}{introduced} the development of AVs and basics of deep learning methods, as well as \textcolor{black}{summarized} recent research on theories and applications of deep learning for AVs. However, they \textcolor{black}{aimed} to identify challenges and solutions in learning algorithms and overview from the vehicle perspective. A summary or discussion from the system perspective, such as the impact of mMP on traffic congestion, has not been presented. Similar conclusions can be drawn from the reviews by\citep{schwarting2018planning,yurtsever2019survey}. To the best of our knowledge, the only work that overview learning-based AV control methods from artificial intelligence (AI) into the field of transportation engineering is \cite{di2020survey}. Nonetheless, the survey primarily was focused on how to deal with interactions between AVs and human-driven vehicles (HDVs), especially in academia works.

Compared to the existing review papers on AV control, the aim of this study is to provide a comprehensive outlook to consolidate the existing knowledge base of upcoming mMP of AVs and their impacts on traffic congestion. 
Specifically, this review paper aims to answer the following questions: 
\begin{itemize}
	\item Data: whether existing self-driving datasets contain congested scenarios? Do they include the necessary features/labels to train a congestion-mitigating mMP? 
	\item Learning method: what are the potential strengths and weaknesses of the typical learning methods in terms of their impact on traffic congestion? 
	\item Domain knowledge: how could the traffic flow expert knowledge help the AI community build the congestion-mitigating mMP?
\end{itemize}

To this end, the paper is organized as follows: Section 2 \textcolor{black}{introduces} available open datasets for AV development; Section 3 summarizes learning methods for AV control; Section 4 discusses the major limitations and challenges arising from these previous works; Section 5 proposes how to utilize traffic domain knowledge to leverage current mMP, and Section 6 presents the discussion and outlook based on this review work.

\section{Datasets for mMP }
\begin{figure}[htbp]
    \includegraphics[width=1.0\linewidth]{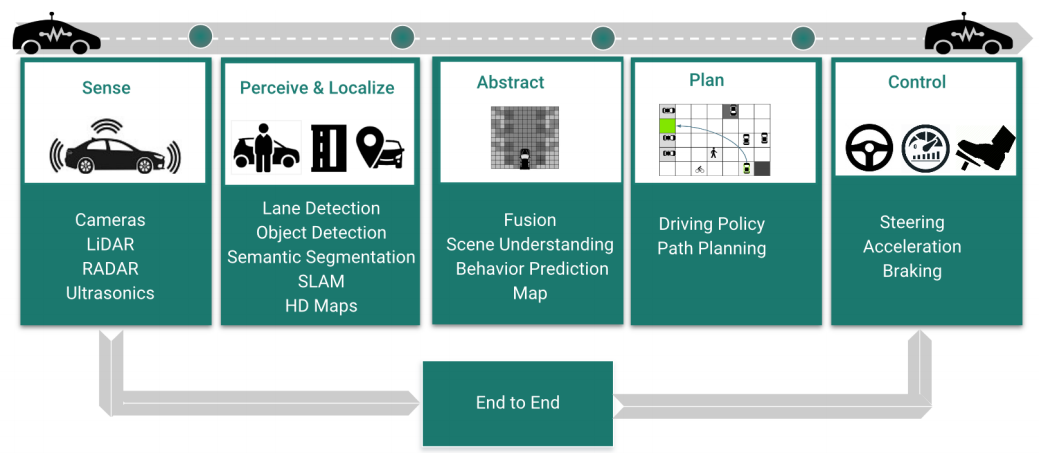}
    \centering
    \caption{Fixed modules in a modern autonomous driving systems. Source: \citep{talpaert2019exploring}}
    \label{fig:module}
\end{figure}

 A typical framework in modern autonomous driving systems is shown in Fig \ref{fig:module}. Among those pillars, the mMP in our paper falls into the driving policy/path planning module. Following the pipeline, we review the related components of training data, model input and output, as well as the learning methods for mMP.  

\subsection{Available open datasets} 
Two recent studies \citep{yin2017use,kang2019test} provided good reviews of the existing open datasets, which covered data scale, contents (camera or lidar, object annotation), road scenarios (urban streets or highway), weather conditions and test vehicle type. Most of current open datasets are designed to assist computer vision development, even leaving out the some required information (e.g. acceleration, trajectory data) to mimic human driving. From the perspective of traffic congestion, Table \ref{datasets} \textcolor{black}{summarizes} the datasets including the position information that is necessary for learning mMP. We also show specific concerns to the driving scenarios and traffic conditions which are certainly related to mMP model.

\begin{table*}[ht]
\caption{Open datasets and simulators for training autonomous driving systems}
\label{datasets}
\begin{threeparttable}
\begin{tabularx}{\textwidth}{lllllX} 
 \hline\noalign{\smallskip}
    \multirow{2}{*}{}{\textbf{Dataset} }   & 
    \multicolumn{2}{l} {\textbf{Data Type}} &
    \multirow{2}{*}{}{\textbf{Scenarios \tnote{1}} }& 
    \multirow{2}{*}{}{\textbf{Traffic } }& 
    \multirow{2}{*}{}{\textbf{Highlights and comments} }                   \\ \cmidrule(lr){2-3}
    &\textbf{T \tnote{2}} &\textbf{C \tnote{3}}\\
\noalign{\smallskip}\hline\noalign{\smallskip}
Waymo \citep{sun2019scalability}                                                    & Y                     & Y      & U                       &  Light                  & Kinematics derived from speed                      \\ \hline
Apolloscape \citep{huang2018apolloscape} & Y                     & Y      & U                       &                             Light and dense                         & Cover different traffic densities                                                              \\ \hline
KITTI \citep{geiger2012we}               & Y                     & Y      & H, U                       & Light and dense & The first AV dataset, mainly used for vision                                                                 \\ \hline
BDDV \citep{xu2017end}                   & Y                  & Y      & H, U  &  Light and dense                   &    A large-scale diverse driving video dataset with comprehensive annotations                                                                                            \\ \hline
Udacity \citep{Udacity}                  & Y                     & Y      & U                      &    Light                   & Driven by ACC, in 2016                                                                         \\ \hline
nuScenes \citep{caesar2019nuscenes}      & Y                     & Y      & H, U               &  Dense                                                & Boston and Singapore, including congestion                      \\ \hline
Ford \citep{Agarwal2020FordMS}           & Y                     & Y      & H, U                        &  Intermediate                                              & Including car-following in congestion                                  \\ \hline
Argoverse \citep{Argoverse2019}          & Y                     & Y      & H, U                       &   Intermediate                                       & Annotation and labels included in the video                            \\ \hline
NGSIM                                                     & Y                     & N   &  H, U                             &    Light and dense     & Mostly used by CF model studies                                                                     \\ \hline
Comma.ai \citep{1812.05752}                                                  & Y                     & Y      &     H                                 &    Intermediate                                                  & Driven by ACC and human drivers                                                                \\ \hline
Brain4Cars  \citep{Jain2016Brain4CarsCT}                                              & Y                     & Y      & H, U         &                           Unclear                           & Behavioral Label                                                                               \\ \hline
CityScapes       \citep{Cordts2016Cityscapes}                                    &       Y               &   Y     & U              &               Dense                                       &       Diverse real-world driving scenes with high-quality annotation                                                                                     \\ \hline
Oxford RobotCar \citep{Barnes2019TheOR}                                        & Y                     & Y      & U    &         Light and dense                                             & Diverse traffic conditions for the whole year in Oxford, UK \\ \hline
UAH \citep{Romera2016NeedDF}                                                      & Y                     & Y      & H                             &        Intermediate                                             & Driving behavior analysis with IOS App                                                             \\ \hline
HighD \citep{Krajewski2018TheHD} & Y & N & H & Light and dense & High-resolution drone data with extracted features\\ \hline
L3Pilot \citep{Hiller2020TheLD} & Y & Y & H, U & Light and dense & The first comprehensive test of OEM autonomous driving systems in EU \\ \hline
\textcolor{black}{ACC data \cite{gunter2020commercially,li2021car, makridis2020openacc}} & Y & N & H & Intermediate & Trajectories of recent ACC car models \\ \hline
\end{tabularx}
\begin{tablenotes}
\item[1] H: Highway; U:Urban.
\item[2] T: Trajectory data.
\item[3] C: Camera data.
\end{tablenotes}
\end{threeparttable}
\end{table*}

Among currently existing dataset, nuScenes \citep{caesar2019nuscenes} and HighD \cite{Krajewski2018TheHD} have shown some consideration of congestion. The nuScenes dataset was collected data from Boston and Singapore, two cities that are known for their dense traffic and highly challenging driving (242km travelled at an average of 16km/h). The HighD dataset was recorded at six different locations near Cologne, Germany. However, we are not aware of any studies that have used nuScenes or HighD to train autonomous driving system. Recently, more AV companies from the industry like Waymo and Lyft have released some open datasets. Waymo's dataset \citep{sun2019scalability} does not provide direct information of trajectory, one needs to derive it using kinamatics information. Lyft's dataset \citep{lyft2019} does not cover congestion scenarios. Tesla has not revealed any plan to publish their dataset yet, but we conjecture their large deployment of vehicle fleets would be highly possible to gain enough congestion data. Remarkably, the L3Pilot dataset will record the autonomous driving behavior and the trajectories of 13 OEM autonomous driving systems, which includes 1000 drivers and 100 cars in various driving conditions (i.e., different weather and traffic conditions) across 10 contries in Europe. The comprehensive coverage and enriched features of L3Pilot dataset can significantly enhance the research of autonomous driving. However, the L3Pilot autonomous project is still ongoing and the corresponding dataset is still not available to the public yet. Thus, the current overall situation indicates the lack of congestion consideration in both academia and industry. Next-Generation Simulation (NGSIM), an open dataset consisted of 2D trajectories, has been widely used CF studies for decades. Different from those datasets from the AV industry, the traffic density in NGSIM often varies significantly and covers both full states from free-flow to traffic jams. It also exhibits a high degree of vehicle interaction near traffic bottlenecks like on-ramp or off-ramps. The diversity of driving scenarios and the interaction among vehicles make NGSIM especially valuable for learning driving behaviors under congestion. However, it does not provide any image or lidar data compatible with sensors for AVs. \textcolor{black}{Moreover, the OpenACC dataset \cite{makridis2020openacc} provides the highway trajectory data of multiple vehicles driven by different commercial ACC systems. However, similar to the NGSIM dataset, the OpenACC dataset does not provide any video data or vehicle sensor recordings that be can leveraged in end-to-end mMP.}

A general issue in those open datasets is that it is unclear whether those miles were driven by human drivers, traditional ACC controllers or new mMP models. It becomes a major limitation when researchers tries to reverse-engineer those current mMP models or simply use the data for training. It might also explain why the applications of those open datasets to transportation studies are still very limited. 
\textcolor{black}{ Overall, the current datasets from the self-driving industry are very limited for analyzing the impact of mMP method on traffic congestion. While more and more commercial ACC products are expected to be equipped with mMP in the future, it would be beneficial for research purposes that car companies share driver data.}

\subsection{Simulator datasets}

While the data from real world is costly to collect, hi-fidelity driving simulators have also been developed to train AVs. CARLA \citep{dosovitskiy2017carla} and TORCS \citep{wymann2000torcs} might be the most popular open-source simulators for autonomous driving research. Related studies based on those simulators can be found in \cite{chen2015deepdriving,panwai2007neural,codevilla2018end,mirowski2018learning,tan2018autonomous}. CARLA can define diverse sensor suites and is also able to generate congested traffic scenarios. A specific method of transferring driving policies from simulations to real world was shown in \cite{muller2018driving}. Note that those simulators also make reinforcement learning (RL) method become feasible by providing an interactive environment for agents to learn.

Both academia and industry have been using simulator datasets to test AV software and hardware. For example, in academia, developers from CMU and MIT used TROCS and Talos simulators respectively, to test their algorithms in simulation before porting them to the vehicle for practical road test \citep{wei2009robust, leonard2008perception}. Recently researchers used simulated LiDAR data to develop and test algorithms for autonomous ground vehicle off-road navigation using MSU autonomous vehicle simulator \citep{dabbiru2020lidar}. \textcolor{black}{To efficiently and effectively supply the critical events and corner cases for the evaluations of AVs, \cite{feng2021intelligent} leverages reinforcement learning algorithms to generate naturalistic adversarial critical events in CARLA to test the safety performance of AVs.} 
In industry, simulator datasets were used by car manufacturers not only to eliminate modeling errors and validate control systems for AVs \citep{short2008assessment}, but also to evaluate the powertrain performance and the analysis of energy consumptions of AVs \citep{zhao2019virtual, cantas2019development}. Waymo \citep{kehrer2018framework, zhao2019virtual} and Uber \citep{Uber2019} developed simulator platforms to generate realistic scenarios from their real-world datasets to improve the safety and performance of AVs. 

\textcolor{black}{For the impact on traffic congestion, similar simulation-based methods can be adopted to generate more driving scenarios related to the traffic efficiency, besides the safety-oriented experiments. However, even though the simulation-based method is efficient in generating supplementary data, simulating realistic behavior of human drivers in a complex traffic environment remains a difficult task, because surrogate models used in simulation will inevitably induce model bias and over-simplified behaviors. The simulation environment constructed with such a surrogate model can lead to undesired and biased performance measure of AVs. Alternatively, we could use a simple CF model known to be string stable to train a string stable mMP.}
Despite the potential benefits of simulator datasets, studies incorporating them to develop a congestion-mitigation mMP model has not been reported. Some studies from transportation research domain might be close \citep{wu2017flow,kreidieh2018dissipating}, which uses a simple traffic simulator to train a single AV to stabilize the mixed traffic. However, we haven't seen the studies using more high-fidelity driving simulator data to investigate string-stable mMP models. 

\section{Learning method}

\subsection{Behavior cloning}
A simple yet effective learning method for mMP is to directly map model inputs to outputs, which can be represented as a function mapping from the input features $s$ (e.g., video frames, figure annotations, kinematic information of ambient vehicles, etc.) to the output action space $a$ (e.g., vehicle speed, acceleration, and steering angle, etc): $F(s)\rightarrow a $. This method is referred as behavior cloning (BC), a subset of imitation learning. The classical framework of BC methods for mMP  can be  classified into three categories: end-to-end learning, mid-level learning, and mixed (hybrid) learning approach. These methods are specifically discussed below:

i) End-to-end mMP. The end-to-end learning approach behaves similar to a black-box, which takes in the raw video data and output the longitudinal vehicle control command (e.g., speed, acceleration, throttle response). Even though the end-to-end approach preserves the advantages of self-optimizing and requiring less manual efforts in implementation, it does confront difficulties and challenges in capturing and processing crucial features from raw video frames. Specifically, the video data in traffic congestion would contain multiple clusters and pose great difficulty to image processing and feature extraction. In addition, the congestion data may contain undesired noise or become excessively random for neural networks to learn, which might trigger under-fitting or over-fitting issues. The strategies of existing literature solely rely on two categories of neural networks: convolutional neural networks (CNN) and recurrent neural networks (RNN). For instances, \citep{kim2017interpretable}, \cite{bojarski2016end}, \cite{Chen2017LaneKeep}, and \cite{sharma2019lateral} utilized deep CNNs concatenating with multiple fully connected layers to predict the vehicle steering wheel angles, which \textcolor{black}{demonstrated} a decent performance in the real-world driving scenario. Moreover, researchers are also contributing to the vehicle longitudinal command. Considering the spatial-temporal characteristics and the memory impact of vehicle longitudinal trajectories, the Long Short-term Memory (LSTM) or Gated Recurrent Unit (GRU) augmented deep CNN \citep{Eraqi2017, xu2017end, hecker2018end} are applied to artificially forget or remember the historical frame features to improve the accuracy of vehicle longitudinal commands (i.e., speed, acceleration) prediction. 

ii) Mid-level learning. The mid-level learning method is more interpretable compared to end-to-end learning approach due to its explicit hierarchical structure. The first segment of mid-level learning is to extract the useful car-following features (e.g., inter-vehicle spacing, relative speed, lane position, etc) using computer vision algorithms, then the second segment \textcolor{black}{correspondingly retrofits} the car-following model with specific neural network. Remarkably, \cite{zhou2017recurrent} showcased the effectiveness of recurrent neural network (RNN) based car-following model in capturing the traffic oscillation characteristics, which provides an insight on including RNN (e.g., LSTM, GRU) in the deep neural network to retrofit the car-following behavior in congested traffic condition. Moreover, some studies \citep{Deo2018GCN,Lee2019GCN,Su2019GraphCN,Jeon2020GCN} have demonstrated that by arranging the kinematic information of multiple neighbor vehicles in a Laplacian-alike feature matrices or tensors and applying Graph Convolution Nework (GCN) to seize the inter-dependency and social pooling of data, \textcolor{black}{we could} improve the performance in predicting the states of ego vehicles. This phenomenon indicates that features with higher dimension and organized in connected structure might lead to higher accuracy. Under this circumstance, it is also significant to evaluate those hand-crafted features in \textcolor{black}{terms} of the model accuracy and parsimony, such that a trade-off can be achieved between model complexity and accuracy.

iii) Mixed (hybrid) learning approach. As including more useful features in the tensor can boost the prediction performance, some studies have also included another sub-task (e.g., semantic segmentation, image augmentation) to extract those useful features in the training process or incorporated other information (e.g., vehicle kinematic states, ambient traffic information) into the end-to-end learning to improve the model accuracy. For instances, \cite{george2018imitation}, \cite{yang2018end}, \cite{Hsu2018End}, and \cite{Li2020AnLA} pooled the vehicle kinematic information with the features obtained from video frames using concatenating layer to enhance the prediction of steering angle and acceleration. \cite{xu2017end} conducted a semantic segmentation aside of the longitudinal and lateral end-to-end learning, and added the loss function of semantic segmentation to the driving loss function of end-to-end learning to reinforce the prediction accuracy. The researchers pointed out the simultaneous learning of semantic segmentation could outperform both the end-to-end and mid-level learning methods.

Remarkably, BC method has gained wide popularity from industry. In Waymo's research paper \citep{bansal2018chauffeurnet}, they reported that even with 30 million examples and mid-level input and output for motion planning, a pure BC method is not sufficient to train a safe AV. To tackle this, they synthesized more 'corner' cases through adding perturbations to the normal driving data. However, we conjecture it might not lead to much difference since the longitudinal motion planning under normal driving scenarios is not strengthened by 'corner' cases. Although Tesla has not published any official research documents on their motion planning technology, from their investor conference event in April 2019 \citep{TeslaEvent}, we could speculate that Tesla most probably adopts BC method as well, and the supervised learning model is evolving due to the large deployment of vehicle fleets on the roads. \textcolor{black}{Currently, Tesla is adopting a feature-engineering approach rather than an end-to-end method. Evidence can be found from the videos \citep{AP2020} on Autopilot's official website , in which entities such as vehicles, traffic lights, or cones are all labelled and annotated separately.} Moreover, in the ScaleML 2020 \citep{ScaleML2020}, Tesla revealed the neural network architecture applied in the FSD, from which we could realize that Tesla is applying a HydraNet for pooling different neural networks \textcolor{black}{which} conduct different tasks of perceptions and predictions (e.g., labeling, annotation, semantic segmentation, per pixel depth prediction) but share the same backbone. Correspondingly, the HydraNet fuses the information from all cameras and radars to create a bird eye view (BEV) for navigating the vehicle.

\subsection{IRL and GAIL}
Another pipeline of imitation learning is to recover the  implicit reward function of human driving using inverse reinforcement learning (IRL) . IRL defines the cost function of a trajectory $c_\theta$ and maximizes the probability of expert demonstration:
\begin{equation}
    p_{\theta}(\tau) = \frac{1}{Z}\exp(-c_{\theta}(\tau))
\end{equation}
Here $\tau$ is a state-action trajectory, and Z is the integral of $\exp(-c_{\theta(\tau)})$ over all trajectories that are consistent with the environment dynamics \citep{finn2016connection}. The parameters $\theta$ are optimized to maximize the likelihood of the demonstrations. 
If the cost function is learned, one can simply use RL to find the policy that behaves identically to the expert. The first IRL study for autonomous driving originated from \citep{abbeel2004apprenticeship}, which proved that it is possible to 'guess' the cost function for some simple task like highway driving by approximating it with a linear combination of some hand-selected features. Related works can be found in \cite{sadigh2016planning,gonzalez2016high,sharifzadeh2016learning}. However, linear assumption of the reward function will lead to ill-posed problems\textcolor{black}{,} because the probability of expert behaviors can be maximized by many different parameters $\theta$. Thus IRL was extended to maximum-entropy ILR by \citep{ziebart2008maximum}. However, IRL methods are typically computationally expensive in their recovery of an expert cost function and generally requires RL in an inner loop.  

Noticing the immense computational cost in recovery the true reward policy, \citep{ho2016generative} found that human driving behaviors can be mimicked directly using Generative Adversarial Imitation Learning (GAIL) without discovering a cost function first. \textcolor{black}{GAIL trains the self-driving policy $\pi_{\theta}$ to perform expert-like behaviors by rewarding it for “deceiving” a classifier $D_\phi$  that discriminates between the policy and expert state-action pairs.} Suppose driving as a sequential decision-making task following a stochastic policy $\pi_{\theta}(s,a)$, which maps an observed road condition $s$ to a distribution over driving actions $a$. Sample a set of simulated state-action pairs $\chi_{\theta} = {(s_0,a_0), (s_1,a_1)...(s_T,a_T)}$ using parameterized policy $\pi_{\theta}$, and the expert behaivor pairs $\chi_{E}$ from $\pi_{E}$, the GAIL objective is: 
\begin{align}
    \max_{\phi}\min_{\theta}V(\theta,\phi) &= \mathbb{E}_{(s,a)\sim\chi_E}[\log D_\phi(s,a)] \nonumber
    \\ &+ \mathbb{E}_{(s,a)\sim\chi_{\theta}}[1-\log D_\phi(s,a)]
\end{align}

In a recent work \cite{kuefler2017imitating}, GAIL was applied to the task of autonomous driving on highway scenario using NGSIM dataset. The result shows that the recurrent GAIL is surprisingly able to capture many desirable properties consistent with real trajectories. \cite{bhattacharyya2018multi} extended GAIL to multl-agent learning for highly interactive driving cases. Although the methodology of GAIL is sound, we have not seen more follow-up studies from the academic community or industry.

\subsection{Reinforcement learning}
\begin{figure}[!htbp]
    \includegraphics[width=0.9\linewidth]{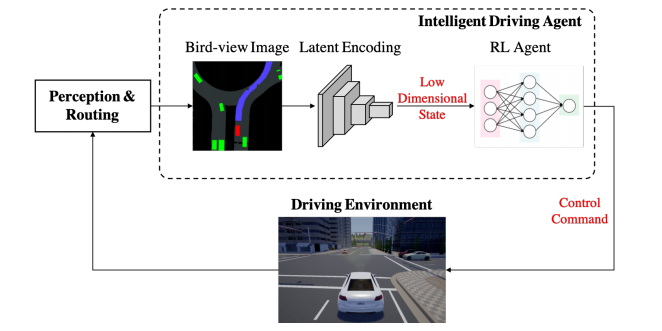}
    \centering
    \caption{Architecture of training autonomous driving in simulation using RL. Source: \cite{chen2019model}}
    \label{RL_frame}
\end{figure}

The the success of imitation learning largely depends on the availability and distribution of labeled data, which are costly to collect. To circumvent this problem, another stream in mMP is working on the non-imitation method, RL, which follows a pipeline shown as Fig.\ref{RL_frame}. Since RL methods need expert-designed reward functions, \textcolor{black}{they can be designed according to the basic driving rules for autonomous driving, such as gaining faster speed and avoiding collisions}. \cite{pan2017virtual} used RL to train \textcolor{black}{an} autonomous driving policy with a pre-defined reward function encouraging higher speed and penalizing crashes. A more recent work \citep{chen2019model} implemented several deep RL methods and showed good driving performance with dense surrounding traffic. \textcolor{black}{
\cite{guo2021hybrid} used RL method to learn the longitudinal motion planning for AVs to reduce the fuel consumption as well as to maintain acceptable travel time.}
\cite{shalev2016safe} applied multi-agent RL in a highly interactive merging case to generate a set of feasible trajectories and then feed a hand-designed cost function to the trajectory planner to select the most smooth and safe trajectory, which makes the longitudinal motion planning no longer a pure BC process. 
DeepTraffic \citep{fridman2018deeptraffic}, a simulation and deep RL environment developed by MIT, has also shown success of RL in navigating AVs through a congested 7-lane highway. Other similar studies based on RL and traffic simulators can be found in \cite{sallab2017deep,kendall2018learning,liang2018cirl}.

It is worth noting that in the RL context, the model input also plays an important role because it directly determines the state space that a RL agent can observe.  \cite{chen2019model} reduced the state complexity through feature representation based on the raw image, which makes the problem more tractable and computationally efficient. 
Despite some studies using RL to stabilize mixed traffic in a loop \citep{stern2018dissipation} or near the merging areas \citep{wu2017flow}, we haven't seen any success in learning a string-stable mMP for single AVs.   

\begin{figure}[!htbp]
    \includegraphics[width=1.0\linewidth]{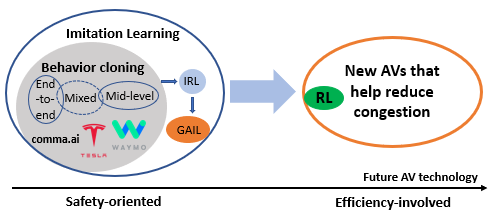}
    \centering
    \caption{Trend of major learning methods for mMP  of AVs}
    \label{trend}
\end{figure}

In summary, we can show the \textcolor{black}{involvement} of the major motion planning methods in Fig \ref{trend}, which depicts the transition from traditional rule-based (RB) methods to the state-of-the-art machine learning (ML) methods. Note that most learning methods fall into the range of BC, and although many alternative learning methods for BC have been proposed in the literature, the leading AV \textcolor{black}{companies} still stick to BC \citep{TeslaEvent,bansal2018chauffeurnet}. Here we do not consider RL as \textcolor{black}{a} BC method since RL does not directly learn from expert demonstration. It does not require large amount of data but a high-fidelity simulator. Also, the performance of RL heavily depends on the human-designed reward functions which governs the training process and resulted policies.

\section{Limitations of mMP}
Based on previous review, in this section we will discuss the current limitations of mMP research regarding its impact on traffic congestion.

\subsection{Systematic lack of training data} 
The datasets that can completely cover regular driving scenarios are still unavailable, let alone the 'corner' cases threatening the robustness of mMP . We found no driving data for  multi-lane highways, on-ramp and off-ramp bottlenecks, or generally congested traffic conditions. Since most neural-network methods cannot generalize well to unseen situations, we believe the incompleteness of datasets might lead to biased or even unpredicatable CF behaviors. Issues of such limitations in biased datasets were also discussed by \cite{codevilla2019exploring}.

\subsection{Incomplete feature representation} 

While perception modules can extract human-interpretable features as model inputs for mMP , we suspect that those hand-selected features may not fully capture all the influencing factors for driving decisions. For example, the specific location information might be totally ignored in model input. From the industry\textcolor{black}{,} no information has been revealed about whether the localization results are incorporated into motion planning. While human drivers respond to different locations with varying driving behaviors, such as the 'relaxation' phenomenon discovered by \cite{laval2008microscopic}, we still do not know whether mMP will react differently in traffic bottlenecks\textcolor{black}{,} such as on-ramp or off-ramps.

\cite{codevilla2018end,sauer2018conditional} conditioned the BC with high-level command input for intersections. The included high-level commands are able to resolve ambiguities in the mapping from single image input to low-level commands (steering and speed). We argue that in highway driving, such ambiguities will also arise between the exiting and non-exiting vehicles \citep{laval2008microscopic}. Thus it would be worthwhile for AVs to incorporate driving intention in motion planning. However, only Tesla \citep{TeslaEvent} has reported a related project to infer the lane change intention of leading vehicles and integrate that for motion planning.

\subsection{Limitations in learning algorithms} 
According to \citep{kuefler2017imitating}, the BC method has been successfully used to produce driving policies for simple scenarios such as car-following on freeways. However, \cite{wheeler2016analysis,lefevre2014comparison} reported different results when applying BC  to nuanced states with little or no experience, showing that BC can only produce accurate predictions up to a few seconds. Their results indicate BC usually demands large amount of training data, and becomes inaccurate when generalized to unseen experiences. Remarkably, as we include LSTM or GRU in the neural network to retrofit the longitudinal command, \textcolor{black}{these} two types of RNN could also face difficulties in transfer learning, posing challenges in generalizing the model. Moreover, the stop-and-go speed profile and the fluctuated and choppy acceleration triggered by congestion appreciably contributes to the difficulties of retrofitting the vehicle longitudinal command, entailing a more intelligent neural network model to capture the fluctuations and discontinuities in the vehicle car-following model during congestion. The poor data distribution generated from driver heterogeneity in congestion \textcolor{black}{also contributes} to the randomness of the model trained by BC, which casts extra doubt on the generalization of a BC model. Thus, BC could significantly suffer from the scarcity of training data and can be biased due to the poor data distribution.

Although IRL and GAIL can circumvent some of the issues with BC method, they still succumb to the pitfall of imitation learning methods. \cite{chen2019model} summarized three major issues with imitation learning: i) it needs to collect a huge amount of expert driving data in real-world and in real time, which can be costly and \textcolor{black}{time-consuming}, ii) it can only learn driving skills that are demonstrated in the dataset. This might lead to serious issues given unseen experience in test process, and iii) since the human driver experts act as the supervision for learning, it is impossible for an imitation learning policy to exceed human-level performance. From the traffic flow perspective, we argue that either BC or other deep imitation learning methods will be cumbersome, \textcolor{black}{especially} with incomplete datasets lacking those important driving scenarios mentioned above. 
According to \cite{gao2018reinforcement}, both BC and IRL algorithms implicitly assume that the demonstrations are complete, meaning that the action for each demonstrated state is fully observable and available. Obviously, this assumption does not hold for the mMP  problem. 

The limitations with imitation learning methods highlights the potential of non-imitation \textcolor{black}{methods} like RL in learning a "better" driving policy to reduce congestion and improve overall traffic efficiency. It is not easy to achieve, though. The major \textcolor{black}{issue} with using RL method is the dependence on a reward function\textcolor{black}{,} which must be hand-crafted based on engineering experience and has to be applicable to all driving scenarios \citep{makantasis2019deep}.  RL methods might cause undesirable driving behaviors by directly transferring their driving policy learned in non-congestion states. Besides, we argue that adopting RL transforms the problem of LLMP from imitating human demonstrations to searching for a policy complying a hand-crafted reward rule. Also, it should be pointed out that RL requires high-fidelity simulation platforms, which must be able to accurately model environment appearance, physics of vehicles and the behavior of other participants \citep{TeslaEvent}. Especially important is the  modeling of vehicle dynamics to accurately represent the effects of gravity, which has been found to be a key factor in reproducing empirical traffic flow instabilities  \citep{laval2004hybrid}.

In summary, RL seems to be our only hope to develop "optimal policies" that could potentially outperform human drivers. Despite the difficulty in designing a good reward function, and the requirement of a more realistic traffic environment, we believe the "trail-and-error" principle in RL is worth borrowing. Note that Tesla seems already working along this direction\textcolor{black}{,} and it is able to use the natural traffic environment to test their algorithms and collect ground-truth data. Again, it remains unknown whether Tesla has considered the congestion impact in their development program.

\section{Traffic domain knowledge}

Overall, the current mMP is devoting most of its efforts to the long-tail safety problem, \textcolor{black}{while its} impact on congestion has been almost completely ignored. Through the above review, we have identified the major limitations in current datasets and learning methods, and now we try to propose some potential future studies which \textcolor{black}{aim} at equipping the learning process with related traffic domain knowledge to fill in the research gap.

Here we \textcolor{black}{summarize} the main intellectual achievements in traditional CF theory which are probably worth-noting for learning approaches to combine. Towards the impact on traffic congestion, the most important human CF properties might include: memory \& prediction, randomness and string stability. 

\subsection{Memory and prediction}
For memory \& prediction, long short-term memory (LSTM), a type of recurrent neural network (RNN), has been adopted by mMP  studies \citep{xu2017end,morton2017simultaneous,bansal2018chauffeurnet} to address the memory impact on future speed choice. Researchers \citep{lefevre2014comparison} conducted a comparative evaluation of parametric and non-parametric approaches for speed prediction during highway driving. The study showed that the CF models can perform well for short-term speed prediction\textcolor{black}{,} but deep neural networks behave better for long-term prediction. To evaluate the relative performance of different learning methods on the same dataset, \cite{kuefler2017imitating} compared GAIL and BC method using the same 2D trajectories from NGSIM. Their work demonstrated that BC has the best short-horizon performance, and GAIL outperforms other methods including CF models for long-horizon task. 

The CF models have realized the merits of introducing memory to improve prediction since long time ago\textcolor{black}{. Studies} also attempted to make some modifications to the traditional CF models basing on its original form. \cite{lee1966generalization} revised the linear GHR model \citep{Chandler1958,gazis1959car} to account for the relative speed over a period of time: $a_n(t) = \int_{0}^{t}M(t-s) \Delta V_{n}(s)ds$, where $M$ is the weight function for the memory impact. \cite{tang2009extended} extended the optimal velocity (OV) model \citep{bando1998analysis} and found that considering human drivers' memory would improve the string stability of traffic flow. Similarly, \citep{zhou2017recurrent} \textcolor{black}{captured} traffic oscillations using a RNN-based CF model, which indicates the memory and prediction can help make informed driving decisions for  smoother traffic. 

It appears that mMP  is able to imitate human driving with such memory and prediction property. For example, AVs will decelerate in advance when realizing a potential decelerating or cut-in behaviors ahead of them. Notably, \cite{TeslaEvent} also mentioned that Tesla can even predict a curving path that cannot be seen by humans due to geometry or limited sight distance. The prediction power of mMP  might outperform human behavior. Also, Tesla demonstrated that their prediction can be used to infer the intention of other vehicles, such as cut-in behaviors which will be incorporated in their motion planning. We conjecture such prediction can improve traffic stability\textcolor{black}{,} because AVs can predict disruptive lane changes and prepare to decelerate first, instead of abrupt deceleration without any prediction. Those studies and new technologies pertinent to memory \& prediction helped to demonstrate the potential of AVs to dampen future traffic congestion.

\subsection{String stability and safety}
The literature has shown that most CF models are unable to replicate string stability consistent with empirical human driving data. These models are all deterministic, including stimulus-response models \citep{gazis1959car}, Optimal velocity models \citep{bando1995dynamical}, IDM model \citep{treiber2000congested} and FVDM model \citep{jiang2001full}), safe-distance model \citep{gipps1981behavioural}, desired-headway model \citep{bullen1982development}, and psycho-physical models \citep{michaels1963perceptual,wiedemann1974simulation}.

\cite{sun2018stability} conducted a comprehensive review on the methods for stability analysis and their applicability to CF models. In this study, they classified the traditional CF models into three categories, which are basic CF models, time-delayed CF models and cooperative CF models based on the assumption of a connected environment. Common methods in the literature for string stability analysis have also been reviewed in detail. \textcolor{black}{However}, those methods applicable for traditional CF models do not apply to mMP  due to its lack of explicit mathematical formulations. 

More importantly, the authors \textcolor{black}{pointed} out some inconsistency between the results using analytical method and numerical simulation, which may result from some of the major assumptions or relaxations: i) since the methods for string analysis are mostly based on linear equations, the non-linear CF models are approximated\textcolor{black}{,} which causes certain numerical errors; ii) the platoon is always assumed to remain in equilibrium before a small perturbation is added when analyzing string stability, which goes against real traffic conditions where different driving regimes need to be considered, and iii) the methods of linear stability analysis is only suitable for small perturbations and the nonlinear effects caused by large perturbations such as hard braking do not apply.  Those studies indicate the string stability of mMP  will be hard to capture due to the non-linear neural networks architectures. Reasonable methods should depend on numerical studies. Therefore, to analyze the string stability of mMP , one has to approximate those proprietary "black boxes" with traditional CF models or a separate neural network, and then conduct numerical simulations for further analysis.

Moreover, safety (collision prevention) is another significant issue in mMP (actually could be weighted the most in autonomous vehicle control design). In the congested traffic, due to the randomness and disturbances induced by human drivers, abrupt braking could be inevitable to guarantee safety, which could consequently jeopardize the string stability performance. Under this circumstance, how will mMP trade off collision-avoidance and the smoothness of traffic in congestion remains to be analyzed and researched on. \textcolor{black}{A} feasible direction could be: making the collision-avoidance as a local-level safety objective while using string stability as a system-level safety objective, and mMP will iteratively optimize these two objectives. Specifically, the local safety objective monitors the immediate safety status of the ego vehicle, preventing collisions with adjacent vehicles during driving tasks. The system-level safety objective could be evaluated as a long-term target, whose focus will be the smoothness (string stability) of traffic\textcolor{black}{. The reason is} that the smooth traffic can alleviate the fluctuations of acceleration and enforce vehicles to operate closer to the equilibrium, which can further prevent collisions in the surrounding traffic. Correspondingly, a specific boundary function \textcolor{black}{needs} to be scrutinizing the safety status during AV operation\textcolor{black}{. Beyond the boundary,} mMP can resort to optimize the system-level performance to alleviate traffic oscillation, while within the boundary, the value of local safety will overwhelm the system-level string stability concern. Therefore, the smoothness of traffic can be an essential criteria of how AVs fit in the traffic in a long term perspective, while collision prevention \textcolor{black}{is} the critical function for AVs to operate safely in a short time span. This is \textcolor{black}{a} significant issue to be carefully balanced such that AVs can scale up and benefit the traffic system.

\subsection{Randomness}

\cite{Laval2014} showed that stochastic errors during the acceleration process are the core of the stop-and-go waves. They developed a parsimonious family of car-following models that are able to reproduce most traffic instabilities, including traffic oscillations and capacity drop, based on stochastic processes to describe drivers' desired accelerations. It was found that this component is crucial for capturing realistic formation and propagation of traffic oscillations. 
This is probably the simplest CF model that captures driver random errors while accelerating and produces realistic traffic oscillations. Follow-up models that incorporate human error have also been formulated within this family  \citep{Xu2019TRR,Kai16} and also for other well-known CF models  \citep{Treiber2017}.

To the best of our knowledge, the stochastic property of mMP  has not been well addressed or used for analyzing traffic congestion.  We argue, however, that it is not advisable to add stochastic components to these methods because it will result in exacerbated traffic oscillations. On the contrary, one should try to minimize this error as much as possible, which should have a positive effect in congestion.

\subsection{Connections between CF models and neural networks}

While most mMP  methods do not show a direct relationship with traditional CF models, it was revealed  that \textcolor{black}{ a mathematical equivalence between mMP  and CF models that can be found under simple settings  \cite{wu2018connections}}. A linear CF model will become interchangeable with a deep neural network given the same input and output. For equivalence in real AV system, \cite{xu2017end}  shows that a mMP  network can be replaced with a traditional CF model given speed and distance extracted from sensor data. We argue that mMP  and CF models are mathematically equivalent if the mid-level methods generate position/distance-based learning affordances (features) as model input for mMP  module. Since CF models adopt design variables of position and speed and output acceleration, the mMP  will boil down to a similar problem which maps the position or speed of surrounding cars to ego-vehicle acceleration. But such equivalence does not apply when the output of mMP  become a predicted trajectory within a few seconds. 

The mathematical connection between mMP and the CF models should result from the approximation power of neural networks, which has been discussed rigorously in literature. \cite{kolmogorov1957representation} proved a general theorem stating that any real-valued continuous function $f$ defined on a n-dimension cube $I^n (n>2)$ can be represented as:
\begin{equation}
    f(x_{1},x_{2}...x_{n})=\sum_{q=1}^{2n+1}\phi_{q}(\sum_{p=1}^n \psi_{pq}(x_{p}))
\end{equation}
where $\psi$ is a continuous and universal one-variable function, and $\phi$ is continuous monotonically increasing functions independent of $f$ . Thanks to Kolmogorov's theorem,  \cite{kuurkova1992kolmogorov} also gave a direct proof of the universal approximation capabilities of perceptron  networks with two hidden layers. Those studies may help to explain why neural networks can successfully replicate CF behaviors of human drivers and longitudinal control methods of AV.

\section{Discussion and outlook}

This survey serves as a preliminary study to investigate the impact of AVs on traffic congestion in the future.  \textcolor{black}{We found mMP is rapidly developing thanks to the efforts from the leading technology companies like Tesla and Comma.ai. Although mMP has not been widely applied yet, most automakers have already equipped enough hardware (sensors) to their latest car models which makes mMP possible in the short future.} Through the review we also found that the AV industry has been mostly focusing on the long tail problem caused by "corner errors" related to safety, while the impact of AVs on traffic efficiency is almost ignored. In detail, none of the existing public datasets provides sufficient data that can be applied on the training \textcolor{black}{of} a congestion-mitigation mMP, and the major learning approach for mMP adopted by  the  industry is still behavior cloning (BC). Albeit some non-imitation methods such as RL are proposed in  the  literature, we have not noticed success in training a congestion-mitigation or string-stable mMP for AVs in the existing literature, let alone the implementation in industry. 

Research is needed to better understand of the characteristics of mMP and their impact on traffic congestion. We suggest the following research directions:

\subsection{Analyzing the impact of AV by approximation and retrofitting}
Since the current AV technologies are sealed as "black boxes", the only way to understand their behavior and impact is to approximate and retrofit AVs using surrogate models. Noticing a certain level of equivalence between CF models and mMP, we can try to approximate the proprietary mMP  by calibrating specific CF models. Similarly, in light of the universal approximation power of neural networks, it is also possible to find surrogate deep neural network (DNN) models for currently unknown mMP models. Therefore, given a trained mMP , we have two different approaches towards understanding its characteristics, either by calibrating a parameterized CF model or training a DNN as approximation. Both of the two methods will pave the way for further studies to analyze the impact of mMP  on safety, string stability in the traffic congestion. 

\subsection{Data enrichment for congestion-oriented research}
Based on our investigation we find there is insufficient data  suitable for researching autonomous driving mMP in congested traffic. Most existing data are biased to emerging autonomous driving task such as object detection or safety issues in corner cases. Thereby, we recommend the industries and academic institutes to put more emphasis on the autonomous (not human-driven) vehicle data collection in congestion, and potentially publish the data for further insights.

\subsection{Incorporating expert knowledge from traffic domains}
For future development of mMP it is advisable that planning agencies create incentives for the AV industry to put more emphasis on the impact of AVs on traffic congestion, rather than only focusing on the long tail problem of "corner errors". Relevant expert knowledge from traffic domains are worth noting, including but not limited to the properties of string stability revealed by traditional CF studies, impact of memory \& prediction, the stochastic accelerations, and the equivalence between CF models and neural networks.

\subsection{Other discussions}
The paper has mainly surveyed and discussed the mMP for AVs, while leaving some other important factors including connectivity, and the cooperation between AV industry and transportation agencies. The authors believe the emerging technology of connectivity also provides great opportunity to benefit the traffic, as more real-time data enable AVs to execute traffic-friendly control algorithms. Additionally, the cooperation between AV industries and transportation agencies is also essential for improving the AV performance in congested traffic, and providing incentives for the smooth transition from human-driven vehicles to AVs.

\section{Author contribution statement}
The authors confirm contribution to the paper as follows: study conception and design: Jorge Laval, Hao Zhou, Srinivas Peeta; dataset survey: Hao Zhou, Wenchao Wu, Anye Zhou, Yu Wang, Zhu Qing; literature review: Hao Zhou, Yu Wang, Anye Zhou, Zhu Qing; draft manuscript preparation: Hao Zhou, Anye Zhou, Jorge Laval, Yu Wang, Zhu Qing, Wenchao Wu. All authors reviewed the results and approved the final version of the manuscript.

\appendix

\section{Appendix. Major AV technology suppliers and customers.}

Here we include the detailed graph showing the relations of major suppliers and customers in AV technology. A full table is shared via the link:
\url{https://wwc20.github.io/AV-technique-suppliers/}

\nolinenumbers

\begin{figure*}[ht]
\begin{subfigure}
    \centering
\includegraphics[angle=90, width = 0.5 \textwidth]{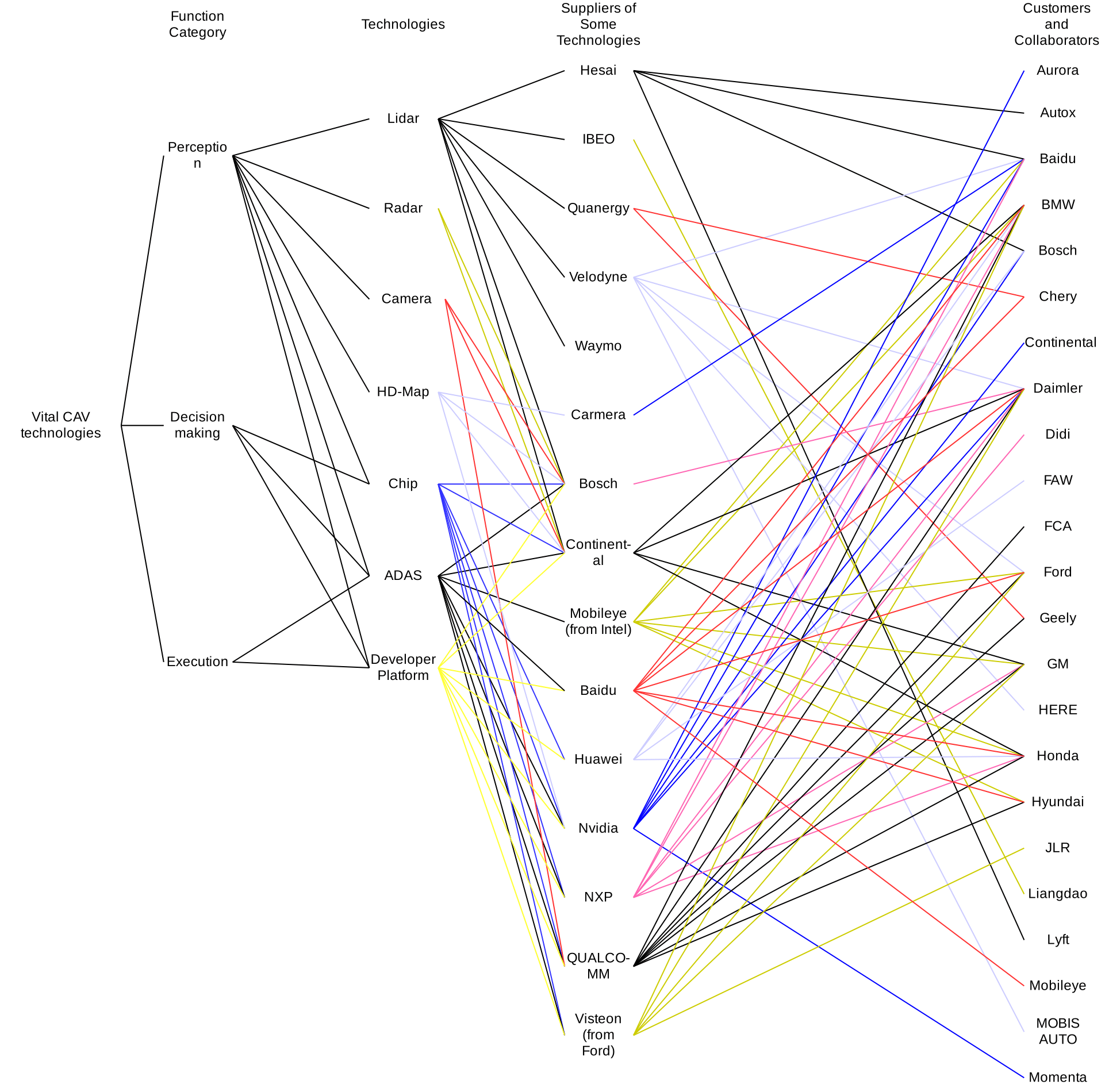}
\end{subfigure}
\begin{subfigure}
\centering
\includegraphics[angle=90, width=0.5\textwidth]{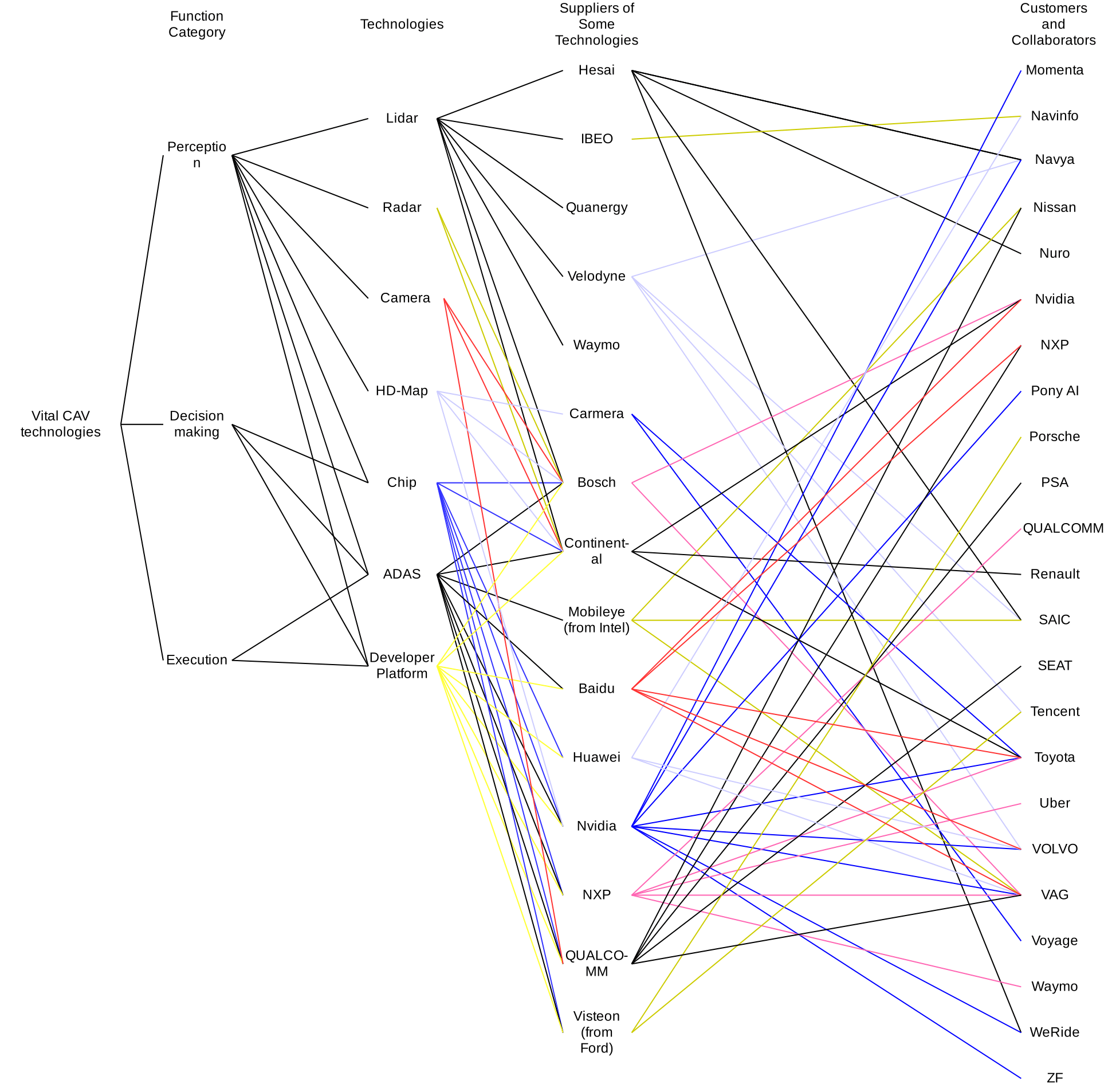}
 \end{subfigure}
 \caption{Suppliers and customers in AV technology. (Source: \cite{Wenchao2020}) }
\end{figure*}
\bibliographystyle{TRR.bst}
\bibliography{bibnew}

\end{document}